\documentclass[fleqn,twoside]{article}

\usepackage{epsfig}
\usepackage{cite}

\usepackage{graphicx}
\usepackage[figuresright]{rotating}

\usepackage{espcrc2}
\usepackage{amsmath}
\usepackage{amssymb}
\usepackage{euscript}

\def\Order#1{${\cal O}(#1)$}

\def\Ordex#1{${\cal O}(#1)_{exp}$}

\def\bbeta{\bar{\beta}}
\def\hbeta{\hat{\beta}}

\newcommand{\Meu}{\EuScript{M}}
\newcommand{\sfac}{\mathfrak{s}}
\title{CEEX Exponentiation in QED%
\thanks{Work partly supported 
  by the European Community's Human Potential
  Programme under contract HPRN-CT-2000-00149 ``Physics at Colliders'',
  by Polish Government grants
  KBN 5P03B09320 %%%<-- Marek
  and 2P03B00122, %(zbw),
  and by NATO grant PST.CLG.977751.
  }}

\author{
  S. Jadach%
  \address{Henryk Niewodniczanski Institute of Nuclear Physics,\\
    ul. Radzikowskiego 152,  31-342 Cracow, Poland 
    }}
      
\begin{document}

\begin{abstract}
  The aim is to summarize briefly on:
  (a) Main features of YFS Coherent Exclusive EXponentiation in QED,
  (b) Examples of recent results relevant for LEP/LC physics program.
\end{abstract}
\maketitle

\vspace{1mm}

This presentation is based mainly on three recent papers
\cite{Jadach:2000ir,Bardin:2001vt,Jadach:2001jx}.
Standard Model calculations for LEP with YFS exponentiation are implemented in 
many Monte Carlo event generators:
for $e^+e^- \to f\bar{f} +n\gamma$, $f=\tau,\mu,d,u,s,c$ process:
(1a) YFS1 (1987-1989) \Ordex{\alpha^1} ISR
      \cite{yfs1:1988},
(1b) YFS2$\in$KORALZ (1989-1990) \Ordex{\alpha^1+h.o.LL} ISR
      \cite{koralz4:1999,yfs2:1990},
(1c) YFS3$\in$KORALZ (1990-1998) \Ordex{\alpha^1+h.o.LL} ISR+FSR
      \cite{koralz4:1994,yfs3-pl:1992},
(1d) KKMC (1998-02) \Ordex{\alpha^2+h.o.LL} ISR+FSR+Interf. ($d\sigma/\sigma = 0.2\%$)
      \cite{Jadach:1999vf};
for the low angle Bhabha process
$e^+e^- \to e^+e^-+n\gamma$ for $\theta < 6^\circ$
the following ones:
(2a) BHLUMI 1.x, (1987-1990) \Ordex{\alpha^1} 
      \cite{Jadach:1989ec},
(2b) BHLUMI 2.x, (1990-1996) \Ordex{\alpha^1+h.o.LL} ($d\sigma/\sigma = 0.07\%$),
      \cite{bhlumi2:1992,bhlumi4:1996};
for the large angle Bhabha $e^+e^- \to e^+e^-+n\gamma$ for $\theta > 6^\circ$
(3) BHWIDE (1994-1998) \Ordex{\alpha^1+h.o.LL}
      \cite{bhwide:1997};
and finally for $WW$ boson production processes
$e^+e^- \to W^+W^-+n\gamma$, $W^\pm \to f\bar{f}$
(4a) KORALW (1994-2001) ISR YFS LL  ($d\sigma/\sigma = 2\%$)
     \cite{koralw:1998,Jadach:2001mp} and
(4b) YFS3WW (1995-2001) YFS expon. + Leading Pole Approx. ($d\sigma/\sigma = 0.4\%$)
     \cite{yfsww:1998,Jadach:2001uu}.

The detailed description of the Coherent Exclusive Exponentiation (CEEX)
can be found in ref.~\cite{Jadach:2000ir}, along with the description
of the older Exclusive Exponentiation (EEX), 
closer to the original Yennie-Frautschi-Suura (YFS) scheme~\cite{yfs:1961}.
What is in a nutshell YFS exponentiation? The main steps in YFS exponentiation are:
\begin{itemize}
\item Reorganization of the perturbative complete \Order{\alpha^\infty} series
      such that IR-finite $\bbeta$ components are isolated (factorization theorem).
\item Truncation of the IR-finite $\bbeta$s to finite \Order{\alpha^n}
      and calculation of them from Feynman diagrams recursively.
\end{itemize}
Apart from disentangling properly ultraviolet divergences from the infrared (IR) ones
YFS deals consistently with the overlapping and sub-leading IR singularities.
To illustrate this point let is show the ``map of IR singularities''
for $2\gamma$ distribution $D_2(k_1,k_2)$ in the plane
of the energies of two real photons:
\begin{center}
  {\epsfig{file=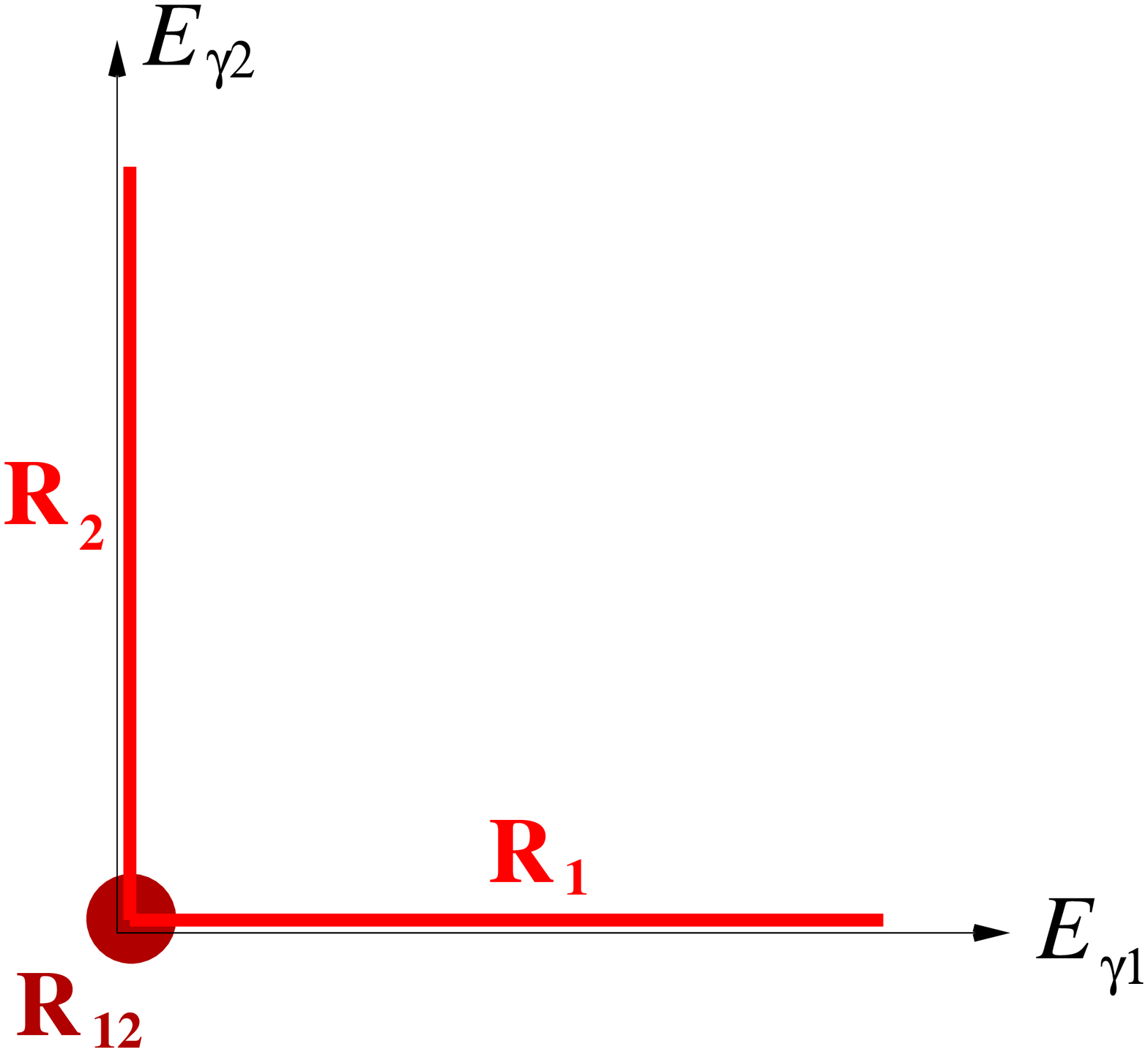,width=40mm,angle=270}}
\end{center}
The point $R_{12}$ is the location of the leading double IR singularity
and the lines $R_1$ and  $R_2$ host non-leading single IR singularities.
The above topological picture translates into a decomposition of the differential
distributions:
$ 
   D_0(p_{f_1},p_{f_2},p_{f_3},p_{f_4})=\bbeta_0(p_{f_1},p_{f_2},p_{f_3},p_{f_4}),
$\\$
   D_1(p_f;{\bf k_1}) = 
        \bbeta_0(p_f) \tilde{S}({\bf k_1}) +\bbeta_1(p_f;k_1), 
$and \\$
   D_2({\bf k_1,k_2}) =
        \bbeta_0 \tilde{S}({\bf k_1})\tilde{S}({\bf k_2}) 
$\\$~~~~
       +\bbeta_1(k_1)\tilde{S}({\bf k_2})+\bbeta_1(k_2)\tilde{S}({\bf k_1})
       +\bbeta_2(k_1,k_2)
$
\\
where typicaly $p_{f_1}+p_{f_2}=p_{f_3}+p_{f_4}$ for $\bbeta_0$ in $D_0$
and $p_{f_1}+p_{f_2}\neq p_{f_3}+p_{f_4}$  for $\bbeta_0$ in $D_1$ and $D_2$.
As we see the IR-finite part $\bbeta_1$ in the decomposition of the 1-photon $D_1$
distribution into IR-leading and subleading parts
is present as an element in the subleading 2-photon $D_2$ distribution.
This nontrivial fact allows one to find out the IR-finite $\bbeta_i$
functions order by order from standard perturbative calculations.
In this particular case the recursive, order-by-order definition of IR-finite $\bbeta$s
looks as follows:
\\
$\bbeta_0(p_{f_1},p_{f_2},p_{f_3},p_{f_4})= D_0(p_{f_1},p_{f_2},p_{f_3},p_{f_4}),
$\\$
\bbeta_1(p_f;k_1) = 
        D_1(p_f;{\bf k_1}) -\bbeta_0(p_f) \tilde{S}({\bf k_1}) 
$\\$
\bbeta_2(k_1,k_2) = D_2({\bf k_1,k_2})
       -\bbeta_0 \tilde{S}({\bf k_1})\tilde{S}({\bf k_2}) 
       -\bbeta_1(k_1)\tilde{S}({\bf k_2})+\bbeta_1(k_2)\tilde{S}({\bf k_1})$
\\
Having at hand the IR-free $\bbeta_i$ the YFS scheme provides 
immediately for the multi-photon
differential distributions all over the phase space and the phase space
is integraded without any approximations using MC method.

Below we show an example of the ISR \Order{\alpha^1} exponentiated multi-photon
fully differential distributions for the CEEX variant of YFS exponentiation
which is implemented in terms of the IR-free $\beta_i$ objects constructed
at the amplitude level, rather than using the spin-summed squared $\bbeta_i$s.
The CEEX we present here very schematically for the process
$e^-(p_1,\lambda_1)+e^+(p_2,\lambda_2)
\to f(q_1,\lambda'_1)+\bar{f}(q_2,\lambda'_2)+\gamma(k_1,\sigma_1)+...\gamma(k_n,\sigma_n)$:
\\
\small{
$
\sigma = \sum\limits_{n=0}^\infty\;
          \int\limits_{m_\gamma} d\Phi_{n+2}\!\!\!\!
          \sum\limits_{\lambda,\sigma_1,...,\sigma_n}\!\!\!\!
          |e^{\alpha B({m_\gamma})} 
          \Meu^{\lambda}_{n,\sigma_1,...,\sigma_n}(k_1,...,k_n)|^2
$},
\\
$\Meu_{0}^{\lambda} = \hbeta_0^\lambda$,\quad $\lambda$=fermion helicities,
\\
$\Meu^\lambda_{1,\sigma_1}(k_1) 
          = \hbeta^\lambda_0 \sfac_{\sigma_1}(k_1) 
          +\hbeta^\lambda_{1,\sigma_1}(k_1)$
\\
$\Meu^\lambda_{2,\sigma_1,\sigma_2}(k_1,k_2) 
          = \hbeta^\lambda_0 \sfac_{\sigma_1}(k_1) \sfac_{\sigma_2}(k_2)
$\\$~~~~~~~~~~~~~~
          +\hbeta^\lambda_{1,\sigma_1}(k_1)\sfac_{\sigma_2}(k_2)
          +\hbeta^\lambda_{1,\sigma_2}(k_2)\sfac_{\sigma_1}(k_1)$
\\
\small{
$\Meu^\lambda_{n,\sigma_1,...\sigma_n}(k_1,k_2,...k_n) 
    = \hbeta^\lambda_0 \sfac_{\sigma_1}(k_1) \sfac_{\sigma_2}(k_2)...\sfac_{\sigma_n}(k_n)
$\\$~~~~~~~~~~~~~~
     +\hbeta^\lambda_{1,\sigma_1}(k_1) \sfac_{\sigma_2}(k_2)\dots\sfac_{\sigma_n}(k_n)
$\\$~~~~~~~~~~~~~~
     +\sfac_{\sigma_1}(k_1) \hbeta^\lambda_{1,\sigma_2}(k_2)\dots\sfac_{\sigma_n}(k_n)\dots
$\\$~~~~~~~~~~
     \dots+\sfac_{\sigma_1}(k_1) \sfac_{\sigma_2}(k_2)...  \hbeta^\lambda_{1,\sigma_2}(k_2)
$}
The \Order{\alpha^1} IR-finite building blocks are:\\
$\hbeta^\lambda_0 = \big(e^{-\alpha B_4} 
      \Meu^{\rm Born+Virt.}_{\lambda}\big)\big|_{{\cal O}(\alpha^1)},$\\
$\hbeta^{\lambda}_{1,\sigma}(k)=\Meu^\lambda_{1,\sigma}(k) - \hbeta^\lambda_0 \sfac_{\sigma}(k)$
\\
It is to be stressed that everything here was done in terms of the spin $\Meu$-amplitudes!
Multiphoton differential distributions are positive by construction (which is not the
case for EEX).
In KKMC the above CEEX is implemented up to \Order{\alpha^2} for ISR and FSR%
\footnote{Multiperipheral contributions and non-IR ISR$\otimes$FSR interferences
    are still incomplete at the \Order{\alpha^2} level.
    They are unimportant for most of LEP observables.}.

\begin{center}
  {\epsfig{file=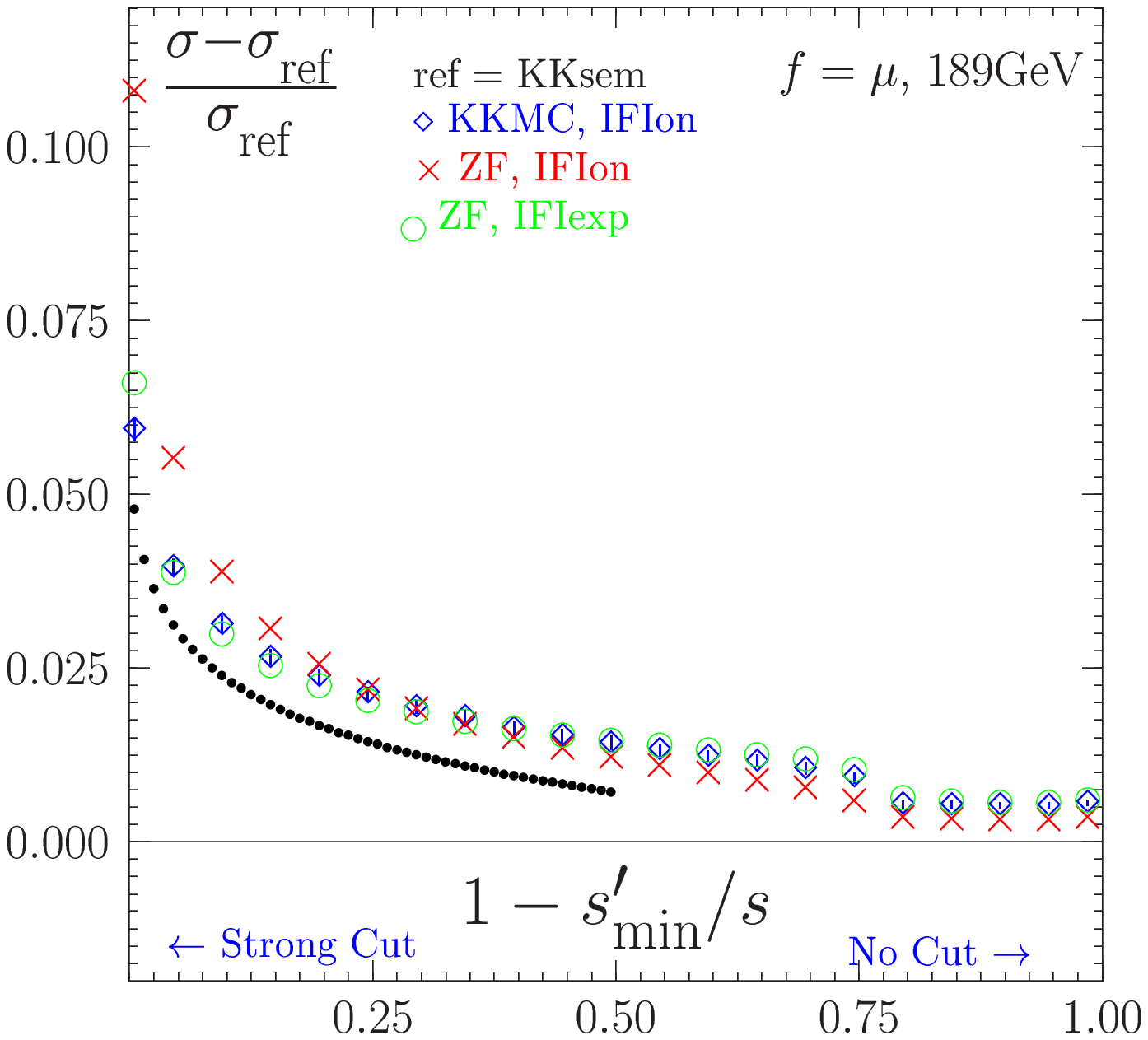,width=60mm}}
\end{center}
The example of the CEEX aplication is shown
in the figure above from ref.~\cite{Kobel:2000aw}
where we present the ``state of art'' comparison of
KK MC vs. Zfitter for ISR+FSR, including IFI=ISR$\otimes$FSR.
We conclude that the precision $d\sigma/\sigma = 0.2\%$ was reached, which is enough
for LEP2 data analysis.
Similar high precision was obtained for the forward-backward asymmetry~\cite{Kobel:2000aw}.

Recently, the CEEX matrix element was extended in KKMC to the neutrino channel.
The contributions from following diagrams are now included:
\begin{center}
{\epsfig{file=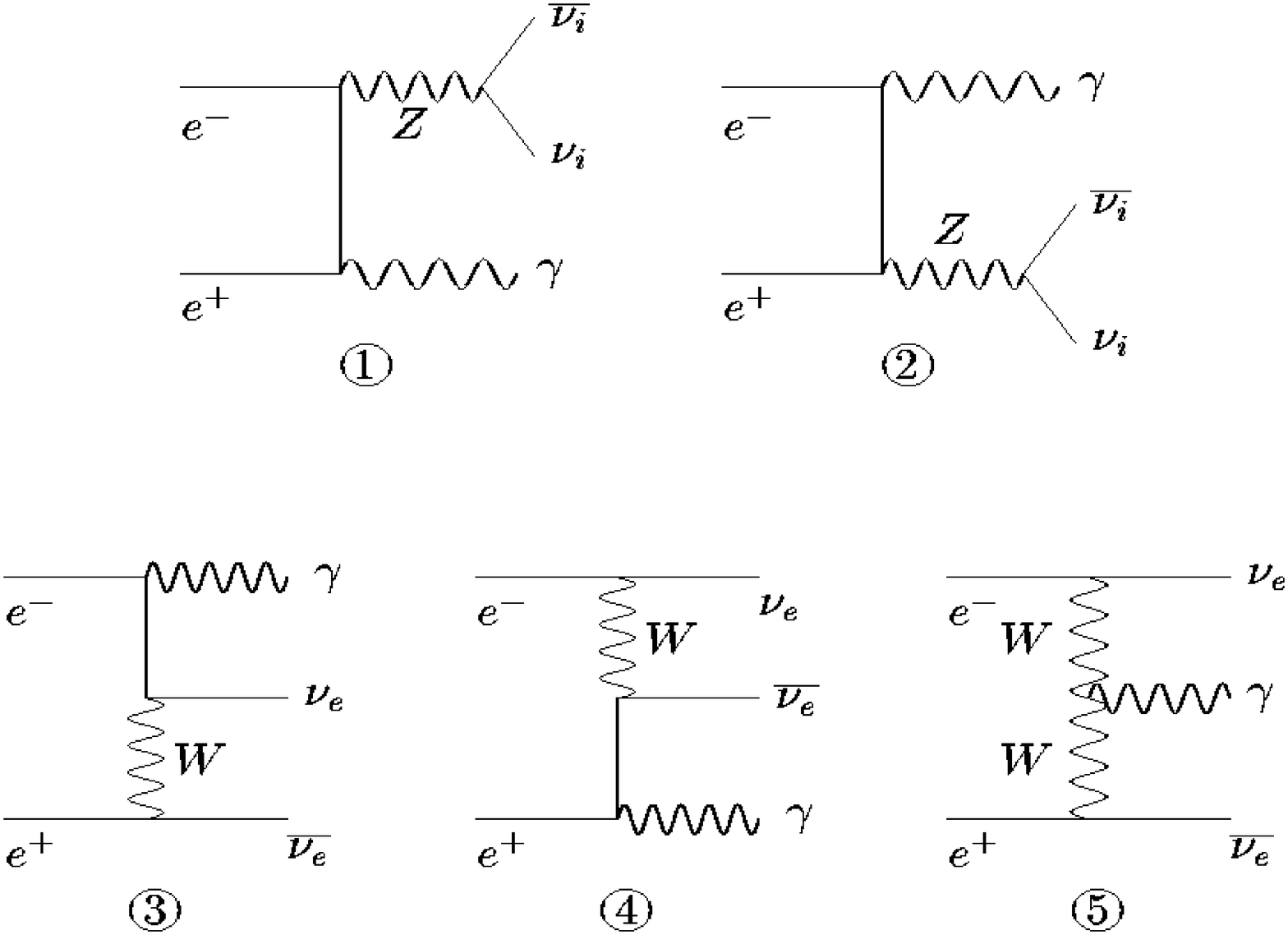,width=70mm,height=40mm}}
\end{center}
and also all double photon emission contributions feature the exact matrix element.
We summarize below on the improvements in KKMC for the neutrino channel.
\begin{itemize}
\item 
  The systematic error is estimated to be
  1.3\%  for $\nu_e \bar \nu_e \gamma$ and 0.8\% for 
  $\nu_\mu \bar \nu_\mu \gamma $ and $ \nu_\tau\bar \nu_\tau \gamma$.
\item 
  For observables with two observed photons we estimate the uncertainty to
  be about 5\%.
\item 
  These new improved results were obtained thanks to
  the inclusion of non-photonic electroweak corrections
  of the {\tt ZFITTER} package
  and due to newly constructed, exact,
  single and double emission photon amplitudes in the KKMC
  for the contribution with the $t$-channel $W$ exchange.
\item 
  The virtual corrections for
  the $W$ exchange are at present introduced in the approximated form.
  The exponentiation scheme CEEX is the same as in the original KKMC program
\end{itemize}

In the first version of KKMC of ref.~\cite{Jadach:1999vf}
the \Order{\alpha^2} next-to-leading contributions were incomplete.
In ref.~\cite{Jadach:2001jx} the \Order{\alpha^2} contribution from the following
diagrams were evaluated
\begin{center}
  {\epsfig{file=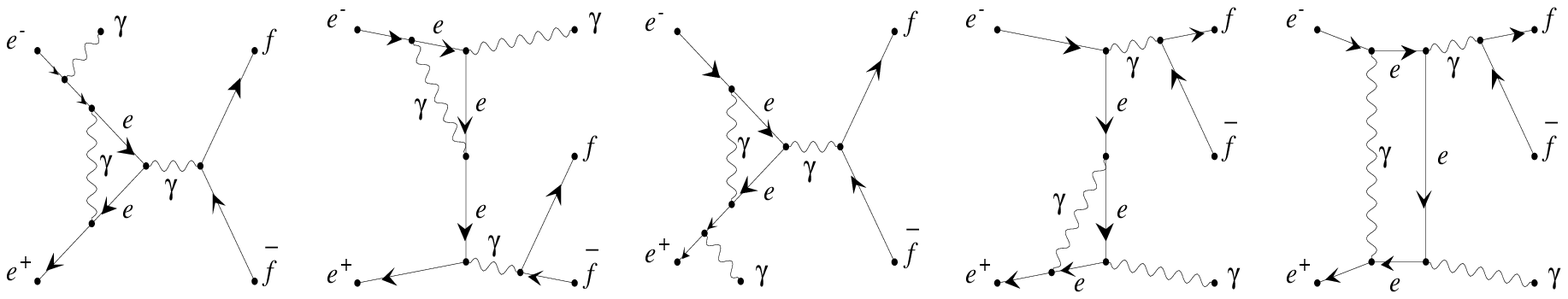,width=75mm}}
\end{center}
and partly included in the development of KKMC --
and the remaining contributions were found to be negligible.
The contributions in questions~\cite{Jadach:2001jx}
are due to one real and one virtual photon emission
and the new numerical results are shown in the following figure
\begin{center}
  {\epsfig{file=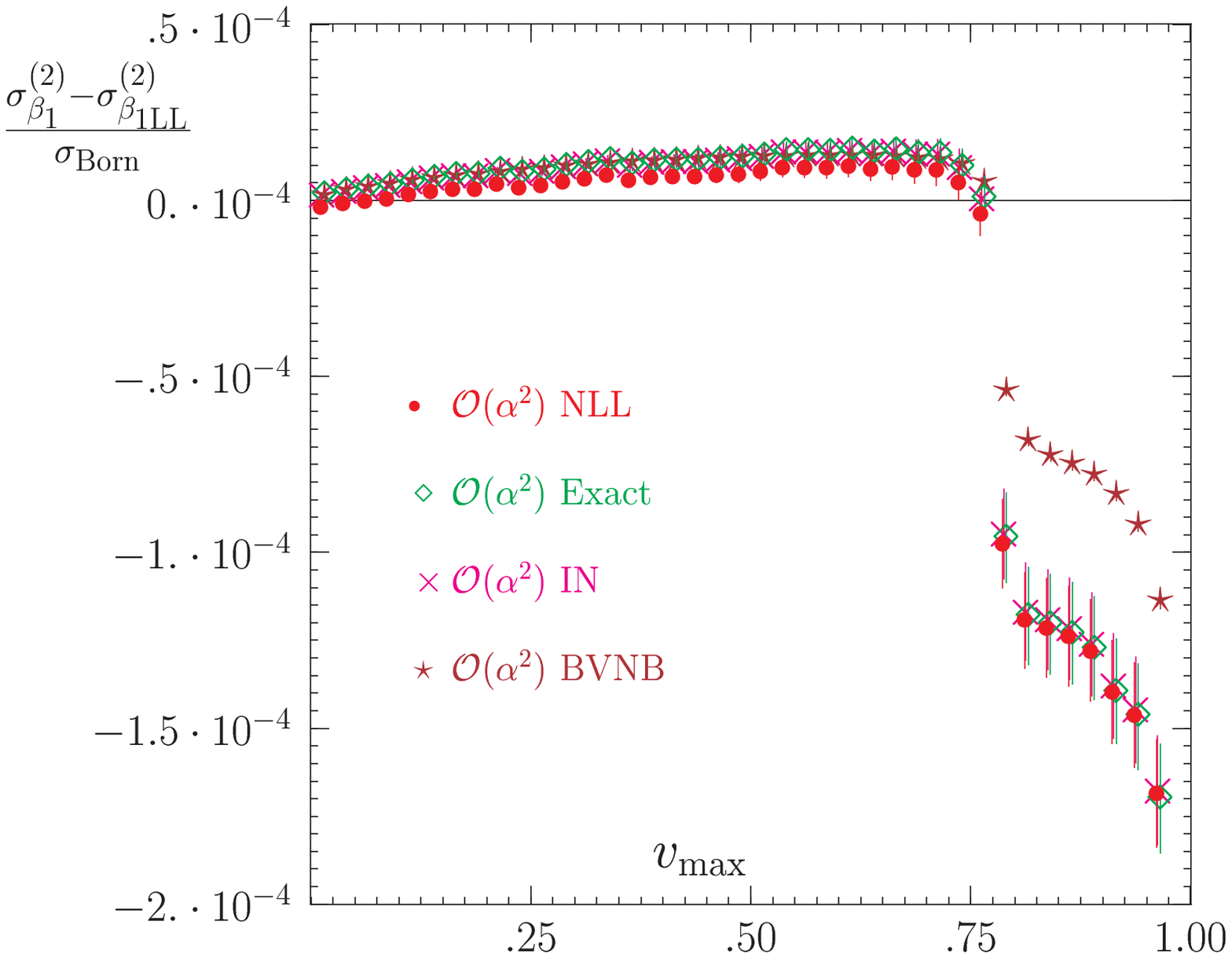,width=50mm}}
\end{center}
where the new calculation is shown to agree with the older calculations
of the refs.~\cite{igarashi:1987,BBVN:1986}.
NB. These contributions were also recently (re)calculated in ref.\cite{Rodrigo:2001kf}
for the purpose of the low energy electron collider experiments.

\vspace{1mm}
\noindent
{\em Conclusions}
The YFS inspired EEX and CEEX schemes are successful examples
of the Monte Carlos based directly on the factorization theorem
(albeit for IR soft case for Abelian QED only).
They Work well in practice: KORALZ, BHLUMI, YWSWW3, BHWIDE, KK MC.
Extension (as far as possible) to all collinear singularities
would be very desirable and practically important!
The KKMC program is extended to the neutrino channel.
Missing fully differential $2f+1\gamma_{virt}+1\gamma_{real}$ distributions
for \Order{\alpha^2} CEEX are now available.

%=============================================================================
%=============================================================================
%=============================================================================
%\bibliographystyle{prsty}
%\bibliographystyle{plain}
%\bibliographystyle{utphys}
%\bibliographystyle{utcaps}
%\bibliographystyle{JHEP-2}
%
%\bibliographystyle{h-elsevier3}
%\bibliography{KK}

\begin{thebibliography}{10}

\bibitem{Jadach:2000ir}
S. Jadach, B.F.L. Ward and Z. Was,
\newblock Phys. Rev. D63 (2001) 113009, hep-ph/0006359.
%%CITATION = HEP-PH 0006359;%%

\bibitem{Bardin:2001vt}
D. Bardin et~al.,
\newblock Eur. Phys. J. C24 (2002) 373, hep-ph/0110371.
%%CITATION = HEP-PH 0110371;%%

\bibitem{Jadach:2001jx}
S. Jadach et~al.,
\newblock Phys. Rev. D65 (2002) 073030, hep-ph/0109279.
%%CITATION = HEP-PH 0109279;%%

\bibitem{yfs1:1988}
S. Jadach and B.F.L. Ward,
\newblock Phys. Rev. D38 (1988) 2897.

\bibitem{koralz4:1999}
S. Jadach, B.F.L. Ward and Z. Was,
\newblock The monte carlo program koralz, for the lepton or quark pair
  production at lep/slc energies: From version 4.0 to version 4.04, 1999,
\newblock hep-ph/9905205, Computer. Phys. Commun. in print.
%%CITATION = HEP-PH 9905205;%%

\bibitem{yfs2:1990}
S. Jadach and B.F.L. Ward,
\newblock Comput. Phys. Commun. 56 (1990) 351.
%%CITATION = CPHCB,56,351;%%

\bibitem{koralz4:1994}
S. Jadach, B.F.L. Ward and Z. W\c{a}s,
\newblock Comput. Phys. Commun. 79 (1994) 503.

\bibitem{yfs3-pl:1992}
S. Jadach and B.F.L. Ward,
\newblock Phys. Lett. B274 (1992) 470.
%%CITATION = PHLTA,B274,470;%%

\bibitem{Jadach:1999vf}
S. Jadach, B.F.L. Ward and Z. Was,
\newblock Comput. Phys. Commun. 130 (2000) 260, hep-ph/9912214.
%%CITATION = HEP-PH 9912214;%%

\bibitem{Jadach:1989ec}
S. Jadach and B.F.L. Ward,
\newblock Phys. Rev. D40 (1989) 3582.
%%CITATION = PHRVA,D40,3582;%%

\bibitem{bhlumi2:1992}
S. Jadach et~al.,
\newblock Comput. Phys. Commun. 70 (1992) 305.
%%CITATION = CPHCB,70,305;%%

\bibitem{bhlumi4:1996}
S. Jadach et~al.,
\newblock Comput. Phys. Commun. 102 (1997) 229.
%%CITATION = CPHCB,102,229;%%

\bibitem{bhwide:1997}
S. Jadach, W. P\l{}aczek and B.F.L. Ward,
\newblock Phys. Lett. B390 (1997) 298,
\newblock also hep-ph/9608412; The Monte Carlo program BHWIDE is available from
  {\tt http://hephp01.phys.utk.edu/pub/BHWIDE}.

\bibitem{koralw:1998}
S. Jadach et~al.,
\newblock Comput. Phys. Commun. 119 (1999) 272, hep-ph/9906277.
%%CITATION = CPHCB,119,272;%%

\bibitem{Jadach:2001mp}
S. Jadach et~al.,
\newblock Comput. Phys. Commun. 140 (2001) 475, hep-ph/0104049.
%%CITATION = HEP-PH 0104049;%%

\bibitem{yfsww:1998}
S. Jadach et~al.,
\newblock Phys. Lett. B417 (1998) 326.

\bibitem{Jadach:2001uu}
S. Jadach et~al.,
\newblock Comput. Phys. Commun. 140 (2001) 432, hep-ph/0103163.
%%CITATION = HEP-PH 0103163;%%

\bibitem{yfs:1961}
D.R. Yennie, S. Frautschi and H. Suura,
\newblock Ann. Phys. (NY) 13 (1961) 379.

\bibitem{Kobel:2000aw}
Two Fermion Working Group, M. Kobel et~al.,
\newblock (2000), hep-ph/0007180.
%%CITATION = HEP-PH 0007180;%%

\bibitem{igarashi:1987}
M. Igarashi and N. Nakazawa,
\newblock Nucl. Phys. B288 (1987) 301.

\bibitem{BBVN:1986}
F. Berends, G. Burgers and W.V. Neerven,
\newblock Phys. Lett. 177 (1986) 1191.

\bibitem{Rodrigo:2001kf}
G. Rodrigo et~al.,
\newblock Eur. Phys. J. C24 (2002) 71, hep-ph/0112184.
%%CITATION = HEP-PH 0112184;%%

\end{thebibliography}

                                     %%%%%%%%%%%%%%%%%%
\end{document}